\documentstyle[psfig]{l-aa}
\begin{document}

\thesaurus{ 08.01.2, 08.02.1, 08.03.3, 08.04.1, 08.12.1, 08.18.1  }

\title {Application of the spectral subtraction technique to the Ca~{\sc ii} H
\& K and H$\epsilon$ lines in a sample of chromospherically active binaries
\thanks{Based on observations made with the Isaac Newton telescope
operated on the island of La Palma by the
Royal Greenwich Observatory at
the Spanish Observatorio del Roque de Los Muchachos of the
 Instituto de Astrof\'{\i}sica de Canarias on behalf of
the Science and Engineering Research Council
of the United Kingdom and the Netherland
Organization for Scientific Research, and with the 2.2 m telescope of
the  Centro Astron\'{o}mico Hispano-Alem\'{a}n of
Calar Alto (Almer\'{\i}a, Spain) operated jointly
by  the Max Planck Institut f\"{u}r Astronomie
(Heidelberg) and the Spanish Comisi\'{o}n Nacional
de Astronom\'{\i}a.} }

\author{D.~Montes
\and E.~De Castro
\and M.J.~Fern\'{a}ndez-Figueroa
\and M.~Cornide}

\offprints{ D.~Montes}

\institute{Departamento de Astrof\'{\i}sica,
Facultad de F\'{\i}sicas,
 Universidad Complutense de Madrid, E-28040 Madrid, Spain
\\  E-mail: dmg@ucmast.fis.ucm.es}

\date{Received ; accepted }

\maketitle
\markboth{D. Montes et al.: Excess Ca~{\sc ii} H \& K and
H$\epsilon$ emissions in chromospherically active binaries}{ }


\begin{abstract}

We present new spectroscopic observations in the Ca~{\sc ii} H \& K line
region for a sample of 28 chromospherically active binary systems
 (RS$~$CVn and BY$~$Dra classes), with different activity levels.
By using the spectral subtraction technique
(subtraction of a synthesized stellar spectrum constructed
from reference stars of similar spectral type and
luminosity class) we obtain the active-chromosphere
contribution to the Ca~{\sc ii} H \& K lines and to the H$\epsilon$ line
when it is present.
We have compared the emission equivalent widths obtained with this technique
with those obtained
by reconstruction of the absorption line
profile below the emission peak(s).

The emissions arising from each
individual star were obtained when it was possible to deblend the
contribution of both components.
The Ca~{\sc ii} line profiles corresponding to different seasons and orbital
phases are analysed in order to determine
 the contribution of each component and to study the chromospheric
 activity variations.


\end{abstract}

\section{Introduction}

The chromospherically active binaries,
RS Canum Venaticorum (RS$~$CVn) systems and BY Draconis
 (BY$~$Dra) stars, defined by Hall (1976)
 and Bopp \& Fekel (1977) respectively, are detached binary systems with cool
components characterized by strong chromospheric, transition region, and
coronal activity. The RS CVn systems have at least one cool evolved component
whereas both components of the BY Dra  binaries are main sequence
 stars (Fekel et al. 1986).

The core emission in the Ca~{\sc ii} H \& K resonance lines are the
most widely used optical indicators of chromospheric activity, their source
functions are collisionally controlled and hence are important diagnostics of
stellar chromospheres.
The chromospheric activity levels can be also inferred by the presence of
emission or filled-in absorption in the core of the H$\alpha$ line and the UV
emission lines.
However, in chromospherically active binaries the excess
emission equivalent widths (EW) of these chromospheric activity indicators
were rarely determined because of
the complication in the analysis due to the
binary nature of these stars.
The reason for this complication is that the
emission of the active component is contaminated by the non
active component which in many cases is the main contributor
to the total observed flux.
In addition, if both components are active their emissions
can be blended.

The calculation of the  active-chromosphere contribution
to the Ca~{\sc ii} H \& K lines
involves the subtraction of a photospheric flux.
The methods normally used to estimate this underlying photospheric
contribution are
the reconstruction of the absorption line profile below the
emission peak(s) by extrapolating the line wings to the line centre
(Fern\'andez-Figueroa et al. 1994 (hereafter FFMCC) and references therein) or
the subtraction of a photospheric contribution obtained from a pure
radiative equilibrium calculation (Linsky et al. 1979).
However, a more suitable method, that eliminates the complications introduced
by the binary nature of these systems in the observed absorption line profile,
is the spectral subtraction technique
(described by Montes et al. (1995a) for the case of the H$\alpha$ line).
This method is based on the subtraction
of a synthesized stellar spectrum constructed
from reference stars of similar spectral type and
luminosity class than the components of the active
system under consideration.

In this paper we study the behaviour of the Ca~{\sc ii} H \& K lines
in a sample of 28 active binary systems of
different activity levels selected from "A Catalog of
Chromospherically Active Binary Stars (second edition)"
(Strassmeier et al. 1993, hereafter CABS).
By using both, the reconstruction of the absorption line profile
and the spectral subtraction technique,
we determine the excess
Ca~{\sc ii} H \& K emissions for each active star.
The dependence of the excess Ca~{\sc ii} H \& K emission EW
on stellar parameters and the relation with
the excess H$\epsilon$ emission EW and
 other activity indicators are studied in a separate paper
(Montes et al. 1995c) in a sample of 73 chromospherically
active binaries where
 the 53 stars analysed by FFMCC have been included.

In $\S$ 2 we give the details of our observations and data reduction
 and we describe the two methods to obtain the
active-chromosphere contribution to the Ca~{\sc ii} H \& K lines.
In $\S$ 3 we describe the individual results of the Ca~{\sc ii} H \& K
line observations of our sample.

\section{Observations and Data Reduction}

Observations of the Ca~{\sc ii} H \& K lines for 28 chromospherically
active binary systems and several inactive stars of similar spectral
types and luminosity classes have been obtained during
two observing runs
in December 1992 and March 1993.
Two different instrumental configurations were used:

1. The Isaac Newton Telescope (INT) at the
Observatorio del Roque de Los Muchachos (La Palma, Spain)
in 1992 December, using the
Intermediate Dispersion Spectrograph (IDS) with grating H2400B, camera 500
and a 1280~x~1180 pixel EEV5 CCD as detector.
The reciprocal  dispersion  achieved  is
 0.18~\AA/pixel
which yields a spectral resolution of 0.36~\AA$\ $ and a useful wavelength
range of 215~\AA$\ $ centered at 3945\AA.

2. The 2.2 m Telescope at
the German Spanish Astronomical Observatory (CAHA) in Calar Alto
(Almer\'{\i}a, Spain) in 1993 March, using a Coud\'{e} spectrograph with the
f/3 camera, grating \#1 and a 1080~x~1030 pixel TEK~\#6 CCD as detector.
The reciprocal  dispersion  achieved  is
 0.21~\AA/pixel
which yields a spectral resolution of 0.42~\AA$\ $ and a useful wavelength
range of 190~\AA$\ $ centered at 3920\AA.

The spectra have been extracted using the standard
reduction procedures
in the MIDAS package (bias subtraction,
flat-field division, and optimal
extraction of the spectrum).
The wavelength calibration was obtained by taking
spectra of a Th-Ar lamp.
Finally, the spectra have been normalized to
the measured flux in 1~\AA$\ $ window centered at 3950.5~\AA.
This reference point at 3950.5~\AA$\ $ is not a real continuum, but it is a
relatively line-free region that could be used as a pseudo-continuum to
normalize all the  Ca~{\sc ii} H \& K spectra and that has been used by
Pasquini et al. (1988) to develop a calibration procedure for converting the
observed line fluxes into absolute surface fluxes.

In this paper we also analyse the spectra of other single inactive stars, used
as reference stars, and other active single stars
and active components of visual binaries taken by us in previous observational
seasons with a similar instrumental configuration (described by FFMCC).

Following our previous paper on Ca~{\sc ii} H \& K emissions in active binaries
(FFMCC) the 28 stars of the sample are arranged in three groups according to
the assigned luminosity class of the active component.
Group 1 contains the systems whose active component is a
main-sequence star (luminosity class V),
group 2 includes the systems whose
active component is an evolved star (luminosity class IV) and
group 3 contains systems whose active component is a giant
or supergiant (luminosity classes III and II)

In Table~1 we list the HD number,  name, spectral type and
luminosity class (T$_{\rm sp}$), the binary nature
(SB), and the adopted  stellar parameters (from CABS)
for the 28 chromospherically active binary systems studied.
In Table~2 we give
the HD number, name, spectral type and luminosity class
(T$_{\rm sp}$), the V-R color, rotational period (P$_{\rm rot}$),
 and Vsin{\it i} (from the Bright Star Catalogue (Hoffleit \& Jaschek 1982),
Noyes et al. 1984; and Donahue 1993)
 of the reference stars used for the
construction of the synthesized spectrum and some other active single stars
and active components of visual binaries which are also studied in this paper.
In the last column A and R mean active and reference star respectively, based
on our Ca~{\sc ii} H \& K observations.

\subsection{Reconstruction of the absorption line  profile}

 In all our papers where the chromospheric emission has been analyzed
(FFMCC, and references therein) the chromospheric
emission fluxes of the  Ca~{\sc ii} H \& K lines, F$_{\rm obs}$(H;K),
were obtained by reconstruction of the absorption
line profile below the emission
peak(s), following the method given by Blanco et al. (1974).
In this method the wing profiles are
extrapolated smoothly toward the line center in order to define the
upper photospheric level.

For the cases in which there is only one emission component,
 the profile reconstruction is easily performed.
However, when the two components of the binary system are active the
observed profile is the result of the emission
from both components and to deblend the two contributions it
is necessary to perform a Gaussian fit of the two emission lines
and the absorption profile.
In some of the examined systems the Balmer H$\epsilon$ line (3970.07 \AA) is
 an emission feature. Since H$\epsilon$ is very near to the
 Ca~{\sc ii} H line (3968.47 \AA) it is also necessary to carry out another
 Gaussian fit to deblend these two emissions.
 The problem is still more complicated
when both components present Ca~{\sc ii} H emission and also one or both
components show H$\epsilon$ in emission, in this case it is necessary to
 perform three or four Gaussian fits.

Following this method, we have determined the
excess Ca~{\sc ii} H \& K emission EW
in the spectra normalized to the pseudo-continuum at 3950.5~\AA.
We have also measured the parameter C(K) defined by FFMCC as:

\[ {\rm C(K)} = \frac{\rm F_{\rm obs}(K)}{\rm F_{\rm T}(K) -
{\rm F}_{\rm obs}(K)} \]
where F$_{\rm T}$(K) is the total Ca~{\sc ii} K emission line flux measured
   above the zero-flux level, whereas F$_{\rm obs}$(K) is the flux measured
 above the reconstructed absorption line profile.
This parameter indicates the relative flux of the emission above the
absorption feature.

\subsection{Spectral subtraction technique}

Another approach to estimate the underlying photospheric contribution is
the subtraction of a synthesized
stellar spectrum constructed from reference stars
of similar spectral type and luminosity class.

This spectral subtraction technique has been extensively used in
the literature for the case of the H$\alpha$ line
(Montes et al. 1994; 1995a; b, and references therein) however for
the case of the Ca~{\sc ii} H \& K lines some attempts have been made:
Catalano (1979) for single active stars, Thatcher \& Robinson (1993)
for a sample of late-G to early-K single and binary stars using only one
reference star of spectral type G6V, Strassmeier (1994b)
for one binary system and finally the work about
composite spectrum subtraction (Griffin et al. 1994 and
references therin).

In this paper we apply to the Ca~{\sc ii} H \& K lines
the spectral subtraction technique described in
detail for the case of the H$\alpha$ line by Montes et al. (1995a).
In this method the synthesized spectra were
constructed from artificially rotationally broadened, radial-velocity
shifted, and weighted spectra of inactive stars (i.e., stars with
negligible Ca~{\sc ii} H \& K emission)
 chosen to match the spectral types of both components of the active
system under consideration. The contribution of each component
 to the total continuum (S$_{\rm H}$ and S$_{\rm C}$)
  has been obtained using the luminosity ratio in the
Ca~{\sc ii} H \& K line region, calculated with the radii and
the Planck functions of the hot and cool components.

By subtracting the synthesized spectrum from the observed one, we
obtain a residual spectrum that contains only the active-chromosphere
contribution to the Ca~{\sc ii} H \& K  lines.
The excess  Ca~{\sc ii} H \& K  emission EW, EW(H;K), is then
defined to be the EW measured in the subtracted spectrum
additively offset to have a continuum of unity.
The contribution to the
profile from each stellar component in the subtracted
spectrum was then
measured by a fit of two Gaussians to the blended profile.
 Finally, the measured EW is
corrected for the contribution of the components to the total
continuum multiplying by a factor ($1/{\rm S}_{\rm C}$)
for the cool component and by
 ($1/{\rm S}_{\rm H}$) for the hot component.

The F$_{\rm S}$(Ca~{\sc ii}~K) flux has been  obtained using
the linear relationship between the absolute surface
flux at 3950~\AA$\ $
 (in erg cm$^{-2}$ s$^{-1}$ \AA$^{-1}$) and the colour
index (V-R) by Pasquini et al. (1988).

\subsection{Measured parameters}

The excess Ca~{\sc ii} H \& K and H$\epsilon$ emissions
EW has been measured following the two above mentioned methods.

In Fig.~\ref{fig:fsfr} we compare the EW obtained from both methods.
As we can see in this figure the EW obtained with both methods are
similar, however the EW measured using the spectral subtraction technique
are always rather larger than the one measured
by reconstruction of the absorption line profile and the difference
tends to be larger for larger EW.
This result was expected since the observed inner wings of the Ca~{\sc ii}
H \& K absorption lines used for reconstruction of the absorption
line profile of the active star, are brighter than the one of the inactive
stars used to the construction of the synthesized spectrum.

The spectral subtraction technique provides better results when the
profile reconstruction is not easily performed, as is the case of systems
that present Ca~{\sc ii} H \& K emissions from both components and systems in
which the absorption line profile have important contribution from both
components.
Furthermore, in some cases, the application of the spectral subtraction
technique allow us to detect a small
H$\epsilon$ emission line that it is not noticeable in the observed spectrum.

The Ca~{\sc ii} H \& K and H$\epsilon$ line parameters,
measured in the observed and
subtracted spectra of the active stars, are given in Tables~3, 4 and 5 for the
groups 1, 2, and 3 respectively.
Column (3) gives the orbital phase ($\varphi$)
for each measured spectrum,
and  in column (4), H and C mean emission belonging to hot and
cool component respectively, and T means that at these phases
the spectral features can not be deblended.
Column (5) gives the weights for the hot and cool component
(S$_{\rm H}$ and S$_{\rm C}$).
In columns (6) to (9) we list the EW and the C parameter for the K and H lines
respectively, obtained by reconstruction of the absorption line profile, and
in columns (10) to (15) we give the EW and the peak emission intensity (I) for
the K, H and H$\epsilon$ lines respectively, measured in he subtracted
spectrum.
Finally, in Table~6 the Ca~{\sc ii} H \& K and H$\epsilon$ line
parameters for the reference stars, the active single stars
and the active components of visual binaries are given.
In columns (3) and (4) we list the core flux, F(1.0\AA), for the
H and K lines, measured as the residual area below
the central 1.0~\AA$\ $ passband.
In the following columns we give for the stars with measurable level of
activity the same parameters that in the last columns of Table~3, 4 and 5.

\section{Individual results}

In the following we describe the Ca~{\sc ii} H \& K spectra of the stars of
this sample and we compare our results, obtained with  the  spectral
 subtraction technique, with the ones reported by other authors.
When observations of these systems in  the  region  of
the H$\alpha$ line are also available (Montes et al. 1995a, b)
we  briefly  describe  the  behaviour  of this emission.

The Ca~{\sc ii} H \& K line profiles of each
chromospherically active binary system
are displayed in Fig.~2 to 29.
 The name of the star, the orbital phase, and the expected
positions of the H \& K  features for the hot (1) and cool (2)
 components are given in each figure.
For each system we plot the observed spectrum (solid-line), the
synthesized spectrum (dashed-line), the subtracted spectrum, additively
offset for better display (dotted line)
 and the Gaussian fit to the subtracted spectrum (dotted-dashed line).
The Ca~{\sc ii} H \& K line profiles for the the active single stars
and the active components of visual binaries are displayed
in Fig.~30 to 42. The orbital ephemeris used for the phase computation are
taken from CABS apart from the cases explicitly mentioned in the text.


\subsection{Group 1 (Active component of luminosity class V)}


\subsubsection{VY Ari (HD 17433)}


This system is a single-lined spectroscopic binary (K3-4 V-IV).
Bopp et al. (1989) found the H$\alpha$ line varying from a
pure emission profile to an absorption profile with occasionally
enhanced H$\alpha$ emission presumably related to flare events.
Our observation in the H$\alpha$ line region (Montes et al. 1995a, b)
shows a weak H$\alpha$ emission superimposed
to the blue wing of a weak absorption.

We have taken one spectrum of this system in Dec-92
at orbital phase 0.17 (Fig.~\ref{fig:hykvyari}).
The spectrum is well matched using a reference  star  of  spectral
type G8IV. The subtracted spectrum
 presents strong Ca~{\sc ii} H \& K
emission lines and the H$\epsilon$ line in emission.


\subsubsection{OU Gem (HD 45088)}


OU Gem is a double-lined spectroscopic binary composed of two
K dwarf stars (K3V/K5V) with strong Ca~{\sc ii} H \& K emission
from both components and filled-in absorption of the
H$\alpha$ line in the cool component (Bopp et al. 1981a, b).

Our observation in the H$\alpha$ line region (Montes et al. 1995a, b)
reveals a strong excess emission but it is not possible to know whether
the emission belongs to the cool component or to both components.

We took one spectrum in the Ca~{\sc ii} H \& K line region in Mar-93 at
orbital phase 0.47 (Fig.~\ref{fig:hykougem}).
In this spectrum we can see strong emission in the Ca~{\sc ii}
H \& K lines and a small H$\epsilon$ emission line. At this orbital
phase the difference in radial velocity between the
components is very small and it is not possible to know whether
the emission arises from the cool component or
rather it is a composite of the two components
 as was observed by Bopp et al. (1981a),
 further taking into account that the spectral type of both
components is similar.
We have constructed a synthesized spectrum with two K1V stars
and a relative contribution of 0.7/0.3.
We have obtained a satisfactory fit between observed and synthesized
spectra in the absorption lines. However in the inner wings of the
Ca~{\sc ii} H \& K absorption lines a significant excess
is present (see Fig.~\ref{fig:hykougem}), which can be interpreted
as a result of noradiative heating of the upper photosphere.


\subsubsection{YY Gem (BD +32 1582)}


YY Gem is a doubled-lined partial eclipsing binary
and one of the most active
flare stars. CABS assigns dM1e/dM1e spectral types to the system.
Our H$\alpha$ observation (Montes et al. 1995a, b) at orbital
phase 0.49 shows a strong emission with contributions from both components.

Our spectrum in the Ca~{\sc ii} K \& K line region at
orbital phase 0.44 (Fig.~\ref{fig:hykyygem}, upper panel)
shows very strong emission from both components.
In the case of the K line the intensity of the emission arising
from the blue component is rather lower than one from the red
component.
In the case of  H line, three
emissions are clearly seen, that is, in order of increasing wavelength:
Ca~{\sc ii}  H of the primary, Ca~{\sc ii} H of the secondary overlapped with
H$\epsilon$ emission of the primary and finally the H$\epsilon$
of the secondary.
In the same spectrum we can also see that  H$\zeta$,
and H$\eta$ Balmer lines from both components appear in emission
(see Fig.~\ref{fig:hykyygem}, lower panel).


\subsubsection{BF Lyn (HD 80715)}


Double-lined spectroscopic binary with spectral types K2V/[dK] and strong
Ca~{\sc ii} H \& K and H$\alpha$ emissions from both components
(Strassmeier et al. 1989;  Barden \&  Nations 1985).

We have taken one spectrum of this system in the Ca~{\sc ii} H \& K line
region at orbital phase 0.21 (Fig.~\ref{fig:hykbflyn}).
This spectrum shows a clear and strong emission in the H \& K lines from both
components with very similar intensities.
The H$\epsilon$ line from the red-shifted component is in emission,
the blue-shifted component
also present H$\epsilon$ emission but is overlapped with the Ca~{\sc ii} H
emission of the red-shifted component.
The synthesized spectrum has been constructed with two K1V stars
and with the same contribution to the total continuum.


\subsubsection{DH Leo (HD 86590)}


This is a triple system whose primary and secondary components
are active.

Our spectra in the H$\alpha$ line region (Montes et al. 1995a, b),
clearly show the H$\alpha$ emission line from the secondary
and an absorption feature from the primary.
The subtracted spectra allow us to obtain the
emission EW of both components.
We found that the hot component presents the
strongest excess H$\alpha$ emission.

Five spectra of this system in the Ca~{\sc ii} H \& K line region are
available. Two observations taken in Feb-88 at
orbital phases 0.32 and 0.55 (FFMCC) and three new observations taken in
Mar-93 at orbital phases 0.07, 0.87 and  0.70 (Fig.~\ref{fig:hykdhleo}).
All these spectra show that  both
components  present Ca~{\sc ii}  H and K emission.
The strongest emission, always centered in the absorption, correspond to the
hot component,  which also  presents  H$\epsilon$  in  emission.
The weak emission blue- or red-shifted depending on the orbital phase belongs
to the cool component.
In the spectra at phases 0.07 and 0.32 the Ca~{\sc ii}  H of the cool
component is overlapped with H$\epsilon$ emission of the hot component.
The bump between the Ca~{\sc ii} emissions in Fig.~\ref{fig:hykdhleo},
(best seen in the second and the fourth panel) could arise, taken into account
data given by Barden (1984), from the third
component of the system.

The synthesized spectra have been constructed with a contribution of each
component to the total continuum of 0.9/0.1, calculated with the
radii and spectral types of the components, and in agreement with the observed
spectrum.

We have not found considerable temporal or orbital phase
variations of the emission.


\subsubsection{AS Dra (HD 107760)}


This is a double-lined spectroscopic binary composed of
two main sequence stars of similar spectral type (G4V/G9V).
Previous observations carried out by us (Fern\'{a}ndez-Figueroa et al. 1986)
showed that the Ca~{\sc ii} H \& K emissions arise from both components and
the hot star was the more active component of the system.

Two new observations of this system taken in Mar-93 at
orbital phases 0.49 and 0.85 confirm that both components are active.
In the spectrum at orbital phase 0.85 (Fig.~\ref{fig:hykasdra}, lower panel)
 the  Ca~{\sc ii} H \& K emission lines
from both components are clearly seen with intensities very similar with the
red component (corresponding to the hot star)
being something larger than the one
arising for the cool star. This difference is a little larger in the case of
the  Ca~{\sc ii} H line, because the H$\epsilon$
emission line of the cool component
is overlapped with the Ca~{\sc ii} H emission of the hot component.
In the observation at orbital phase 0.49
(Fig.~\ref{fig:hykasdra}, upper panel)
is not possible to separate the contribution from each component,
but the subtracted spectrum points out the presence of the
H$\epsilon$ emission line.
A synthesized spectrum has been constructed
with G2V and G8V references stars and with a relative  contribution of each
component of 0.66 and 0.34.


\subsubsection{IL Com (HD 108102)}


A double-lined spectroscopic binary with F8V+F8V spectral types.
Xuefu \& Huisong (1987) and  Eker et al. (1995) reported
strong H$\alpha$ absorption.
Our spectrum of the H$\alpha$ line (Montes et al. 1995a, b) shows
absorption lines from both components and the
subtracted spectrum, points out an excess emission
from both components with very similar EW.

We have observed this system in the Ca~{\sc ii} H \& K line region
 in three different epochs.
Two spectra taken in Jun-85 (Fern\'{a}ndez-Figueroa  et  al.  1986), two
spectra taken in Feb-88 (FFMCC) and a new spectrum
taken in Mar-93 (Fig.~\ref{fig:hykilcom}).
In all spectra we can see Ca~{\sc ii} H and K emissions arising from
 both components.
We have constructed a synthesized spectrum with two F8V stars
and with the same contribution to the total continuum.
For this star, the Doppler shifts used have been obtained
measuring the shifts of several absorption lines in the observed
spectra. The Doppler shifts estimated in this way have never agreed
with those extracted from the corresponding orbital phases (see our comments
on this star in Fern\'{a}ndez-Figueroa et al. 1986).
Using this synthesized spectrum,
we have not found a good fit with the observed spectrum.
This result indicates
a possible misleading spectral classification of this system.

In the subtracted spectra of Feb-88 the blue-shifted component have
a little larger intensity than the red-shifted component, however, in Mar-93
we observed the opposite, and the difference between the two emissions is
lower.


\subsubsection{HD 131511 (HR 5553)}


Single-lined spectroscopic binary classified as K2V and not included in the
first edition of CABS (Strassmeier et al. 1988).
Moderate  Ca~{\sc ii} H \& K emission lines were noted by Heintz (1981),
Basri et al. (1989) and Strassmeier et al. (1990),
who also found variable H$\alpha$ absorption.

We present one observation of this system in the Ca~{\sc ii} H \& K line
 region taken in Mar-93 at orbital phase 0.75 (Fig.~\ref{fig:hykhd131511}).
A moderate emission, centered at the absorption line, is observed.
The spectral subtraction using a K1V  reference  star
 yields a good fit in the absorption spectral lines except in the inner
wing of the Ca~{\sc ii} H \& K absorption lines.


\subsubsection{MS Ser (HD 143313)}


Double-lined spectroscopic binary composed by two K dwarfs stars (K2V/K6V).
This system presents strong Ca~{\sc  ii} H \& K emission
(Strassmeier   et   al.  1990)
and the H$\alpha$ line in emission (Bopp et al. 1981b).

We have taken one spectrum of this system in the Ca~{\sc ii} H \& K line
region in Mar-93 at orbital phase 0.16 (Fig.~\ref{fig:hykmsser}).
This spectrum shows strong emission in the H \& K lines and
the H$\epsilon$ line also in emission.
The observed spectrum exhibits absorption lines from both
components, the blue-shifted lines --which according to
the orbital phase correspond to the hot component--
being deeper than the red-shifted ones.
This is in agreement with the calculated
 contribution of each component to the total continuum (0.88/0.2).
On the other hand the wavelength position of the observed H \& K emission
lines corresponds to the more intense and blue-shifted absorption lines,
therefore, we can conclude that the emission arise from the hot component of
the system.

The spectral subtraction yield a satisfactory fit and
the subtracted spectra allow us to obtain the
emission EW of the H, K and H$\epsilon$ lines.


\subsubsection{KZ And (A and B) (HD 218738) }


KZ And is the component B of the visual pair ADS 16557, and is a double-lined
spectroscopic binary with spectral types dK2/dK2.

This system has been observed at  orbital
phase 0.33 in Jul-89 (FFMCC) and at phase 0.39 in Dec-92
(Fig.~\ref{fig:hykkzand}, upper panel).
Both spectra present  Ca~{\sc ii} H \& K emissions from both components.
In the case of K line the intensities of both emissions are nearly equal and
in the H line region, three
emissions are clearly seen, that is, in order of increasing wavelength:
Ca~{\sc ii}  H of the primary, Ca~{\sc ii} H of the secondary
overlapped with H$\epsilon$
emission of the primary and finally the H$\epsilon$ of the secondary.
We have constructed a synthesized spectrum with two K1V stars with the same
contribution to the observed spectrum.
The excess Ca~{\sc ii}  emission obtained are larger in
 the 1989 observation than in 1992.

We have also observed, (in Dec-92) the component A of the visual pair
ADS 16557 (HD 218739) that is only 15" away and of spectral type G0V.
This spectrum  (Fig.~\ref{fig:hykkzand}, lower panel) shows a small
emission in the Ca~{\sc ii} H \& K lines with an intensity of
I$_{\rm K_{3}}$=0.38.


\subsubsection{KT Peg (HD 222317)}


KT Peg is a double-lined spectroscopic binary with spectral types
G5V and K6V.
CABS only indicates the presence of emission
in the Ca~{\sc ii} H \& K lines in this system and
the behaviour of the H$\alpha$ line is not reported.

One spectrum of this system taken in Dec-92 at orbital phase 0.27
(Fig.~\ref{fig:hykktpeg})
shows  Ca~{\sc ii} H \& K emissions from both components.
The stronger emission, centered at the absorption line, arise from the hot
component, which is the component with the larger contribution to the
continuum (0.9/0.1 according to the radii and T$_{\rm  eff}$ of the
components). The red-shifted and less intense emission corresponds to the cool
component. The computed orbital velocity agrees with
the observed shift.
We have constructed the synthesized spectrum with two reference stars of
spectral types  G2V and K1V
and taking into account the different contribution of both components to the
observed continuum above mentioned.
The emission EW of each component have been determined
in the subtracted spectrum
with two Gaussian fits.


\subsection{Group 2 (Active component of luminosity class IV)}


\subsubsection{UX Ari (HD 21242)}


This double-lined spectroscopic binary (G5V/K0IV)
is a well known RS CVn system and
extensively studied in the literature (Carlos \& Popper 1971;
 Bopp \& Talcott 1978;  Huenemoerder et al. 1989; Raveendran \& Mohin 1995).
Our H$\alpha$ observation (Montes et al. 1995a, b) shows clear
H$\alpha$ emission above the continuum from the cool component
which is superimposed to a weak absorption from the hot component.

Our spectrum in the Ca~{\sc ii} H \& K line region at
phase 0.92 (Fig.~\ref{fig:hykuxari})
shows strong emission from the cool component
and a weak  H$\epsilon$ emission line.
The spectral subtraction using G2V and G8IV as reference
stars  yielded a satisfactory fit if we take a relative contribution of
0.60/0.40 that is very different to that obtained with the spectral types and
radii given in CABS.
This result is in agreement with the observed spectrum since
the emission peak, coming from the cool component,
appear blue-shifted --in agreement with the orbital phase--
 with respect to absorption of the hotter component,
which indicates that the contribution to the observed continuum of the hot
component is larger than that of the cool one.


\subsubsection{HU Vir (HD 106225)}


HU Vir is a K0 subgiant in a close binary system with an unseen secondary
component. It shows strong Ca~{\sc ii} H \& K emission and the H$\alpha$ line
filled by emission (Bidelman  1981; Fekel et  al. 1986).
Strassmeier (1994a), found in this system a big,
cool polar spot from Doppler  imaging
and two hot plages 180$^{0}$ apart, from H$\alpha$ and  Ca~{\sc ii} H \& K
line-profile analysis.

We have observed this system in the Ca~{\sc ii} H \& K line region
in Mar-93 at orbital phase 0.71 (Fig.~\ref{fig:hykhuvir}).
This spectrum shows very strong Ca~{\sc ii} H \& K emission
 and an important H$\epsilon$ emission line superimposed
to the wide  Ca~{\sc  ii} H line. The observed intensity of the emission
(I$_{\rm K_{3}}$=2.4) is similar to the maximum emission of the spectra
reported by Strassmeier (1994a).
A good fit between observed and synthesized
spectra has been obtained. To deblend the
contribution of the  H and  H$\epsilon$ lines
we have used a two-Gaussian fit.


\subsubsection{HD 113816 (BD-04 3419)}


This system is a single-lined spectroscopic binary classified as
K2IV-III and with strong Ca~{\sc  ii} H \& K emission lines
(Buckley et al. 1987; Strassmeier 1994b).
The behaviour of the H$\alpha$ line is not reported in CABS.

We have taken one spectrum of this system in the Ca~{\sc ii} H \& K line
region in Mar-93 at orbital phase 0.68 (Fig.~\ref{fig:hykhd113816}).
In spite of the lower S/N ratio of this spectrum
we can see that this system presents
a strong emission in the H \& K lines with intensities
well above the continuum at 3950 \AA, but lower than reported by
Strassmeier (1994b). We have not found evidence of H$\epsilon$ in emission.
The synthesized spectrum has been constructed with a K1IV reference star.


\subsection{Group 3 (Active component of luminosity class III)}


\subsubsection{5 Cet  (AP Psc, HD 352, HR 14)}


Single-lined spectroscopic binary composed of a K3III star
nearly filling its Roche lobe
and a small hot companion best studied in the ultraviolet
(Eaton \& Barden 1988). Classified as $\approx$F/K1III by Bildelman (1981).
CABS reported strong emission from the cool component, based in the
photometric index S given by Middelkoop  (1982).
The behaviour of the H$\alpha$ line is not reported in CABS.

In our observation in the Ca~{\sc ii} H \& K line
region taken in Dec-92 at orbital phase 0.32 (Fig.~\ref{fig:hyk5cet})
 we can see a weak emission
 centered at the absorption line that contrasts with the strong emission
reported in CABS.

We have not found a  satisfactory fit between the observed spectrum and the
synthesized spectra constructed with a K1III reference star.
This effect could be due to the lower S/N ratio of this
spectrum or to the influence of the hot component to the observed spectrum.


\subsubsection{BD Cet (HD 1833)}


This system is a single-lined spectroscopic binary classified as K1III + F by
Bildelman \&  MacConnell (1973). It presents Ca~{\sc ii} H \& K emission of
Class B from the cool component (Bopp  et  al. 1983; Fekel  et al. 1986) and
the H$\alpha$ line as moderate absorption (Fekel  et al. 1986).

We present here one spectrum of this system in the Ca~{\sc ii} H \& K line
region taken in Dec-92 at orbital phase 0.90 (Fig.~\ref{fig:hykbdcet}).
This spectrum shows strong H \& K emission lines  centered at the absorption
line.
The spectral subtraction  using  a  K1IV  reference  star
yields a good fit.


\subsubsection{$\zeta$ And (34 And, HD 4502, HR 215) }


This single-lined RS CVn binary system is classified as K1~II in CABS,
however, Mewe et al. (1981) give type K1~III.
This system have been previously studied by
us in the Ca~{\sc ii} H \& K and H$\alpha$ lines (FFMCC; Montes et al. 1995a).

Two observations in the  Ca~{\sc ii} H \& K line region
 of this systems are available. One spectrum taken in Oct-1991 at orbital
phase 0.29 (FFMCC) and a new spectrum taken in Dec-1992. at orbital phase
0.69 (Fig.~\ref{fig:hykzand}).
Both spectra show strong H \& K emissions centered at the absorption.
We have obtain a good fit between observed and synthesized spectra.
More observations of the Ca~{\sc ii} H line of this systems can be found in
Shcherbakov et al. (1995).


\subsubsection{$\eta$ And (38 And HD 5516, HR 271)}


Double-lined spectroscopic binary  composed by two G8IV-III
stars, listed in the first edition of CABS (Strassmeier et al. 1988) as having
weak Ca~{\sc  ii} H \& K emission lines (I$_{\rm K}$=3, Wilson 1976).
However it is not included in the
second edition of CABS due to insufficient evidence for
chromospheric activity.
Strassmeier \& Hron (1990) reported a nondetection of Ca~{\sc  ii} H \& K
emission and found a normal H$\alpha$ absorption line.
Xuefu  et al. (1993) neither found any H$\alpha$ emission trace.

However, we have found in one spectrum in the  Ca~{\sc  ii} H \& K line
region taken in Dec-1992 at orbital phase 0.62
(Fig.~\ref{fig:hykeand})
a weak emission in these lines.
This emission is also pointed out by application of the spectral
subtraction technique.


\subsubsection{AY Cet (39 Cet, HD 7672, HR 373)}


AY Cet is a single-line binary composed of a spotted G5III primary and a white
dwarf secondary. It presents strong Ca~{\sc  ii} H \& K
emission lines  (Bopp 1984; Strassmeier et al. 1990), radio
flares (Simon et al. 1985) and a filled in absorption H$\alpha$ line
(Fekel et al. 1986;  Strassmeier  et  al. 1990).

Our observation of this system in the Ca~{\sc ii} H \& K line
region taken in Dec-92 at orbital phase 0.32
(Fig.~\ref{fig:hykaycet})  shows a strong emission
 centered at the absorption line.
The observed emission intensity does not reach
the surrounding continuum and it is
very similar to that found by Bopp (1984).
We have used a G6IV reference star to perform the spectral
 subtraction.


\subsubsection{HD 12545 (XX Tri, BD +34 363) }


HD 12545 is a single-lined spectroscopic binary of spectral type K0III and an
extremely active RS CVn binary.
It possess very strong Ca~{\sc ii} H \& K emission lines and the H$\alpha$
line in emission above the continuum
(Strassmeier et al. 1990; Bopp et  al. 1993)
and a photometric amplitude of 0.6 mag in V, which implies that nearly half
the hemisphere of the star was covered by cool spots (Bopp et  al. 1993).

We have observed this system in the  Ca~{\sc  ii} H \& K line region in
Dec-92 (Fig.~\ref{fig:hykhd12545}).
 The orbital phase calculated with the ephemerides given by
Bopp  et al. (1993) is 0.55.
This spectrum shows very strong H \& K emissions and the H$\epsilon$ line also
in emission. The emission intensity observed
in our spectrum (I$_{\rm  K_{3}}$=4.6) is larger
than the emission intensity observed by Strassmeier et al. (1990) (2-3 times
that of the local continuum).
The synthesized spectrum has been constructed with a reference star of
spectral type K1IV obtaining a satisfactory fit.
We have performed a two-Gaussian fit of the subtracted spectrum in order to
 obtain the emission equivalent widths of the Ca~{\sc  ii} H
and H$\epsilon$ lines.


\subsubsection{6~Tri (TZ~Tri, $\iota$~Tri, HD~13480, HR~642, ADS~1697~A)
\label{sec:6tria} }

This system is the A component of the visual binary ADS~1697, and it is
a double-lined spectroscopic binary classified as F5/K0III.
CABS reported moderate Ca~{\sc ii} H \& K emission and the H$\alpha$ line in
absorption.

Our spectrum in the Ca~{\sc ii} H \& K line region taken in Dec-92 at
 orbital phase 0.87 (Fig.~\ref{fig:hyk6tria})
shows a moderate  H \& K emission from the cool component.
The emission peaks are blue-shifted in agreement with the
computed orbital velocity at orbital phase 0.87.
The synthesized spectrum has been constructed
with F7V and K0III references stars and with a relative  contribution of each
component to the observed continuum of 0.2 and 0.8, however we have not found
a satisfactory fit between observed and synthesized
spectra.

We have also observed, (in Dec-92) the component B of the visual pair ADS 1697
(HD 13480B) that is only 4" distant and is another double-lined
spectroscopic binary, with an orbital period of 2.236 days and
spectral type F6V. This spectrum does not show appreciable
emission in the Ca~{\sc ii} H \& K lines.


\subsubsection{V1149 Ori (HD 37824)}


A single-lined spectroscopic binary classified as K1III + F
by  Bidelman \& MacConnell  (1973),
although Hirshfeld \& Sinnott (1982) list this star as a G5IV.
Our observation in the H$\alpha$ line region (Montes et al. 1995a, b)
reveals a clear excess H$\alpha$ emission.

We have taken two spectra of this system in the
Ca~{\sc ii} H \& K line region
in Mar-93 at orbital phase 0.19 (Fig.~\ref{fig:hykv1149ori}).
These spectra exhibit strong Ca~{\sc ii} H \& K emission lines with an
intensity larger than observed by Bopp (1984).

The synthesized spectrum has been constructed  with a G6IV as reference stars
because we obtained a better fit than using a K giant star.
We have also needed to use a
large rotational velocity than the given in CABS to obtain a good fit between
observed and synthesized spectra.
The subtracted spectrum reveals that
the H$\epsilon$ line is also in emission.


\subsubsection{RZ Cnc (HD 73343)}


This system is a double-lined eclipsing binary classified
as K1III/K3-4III by Popper (1976), the
cooler component filled its Roche lobe and it is in the late
phase of mass transfer (Demircan 1990).
Eker et al. (1995)
found weak H$\alpha$ features from both components and concluded that
both stars have filling in the H$\alpha$ core.

The application of the spectral subtraction to the H$\alpha$
line (Montes et al. 1995a, b) reveals
the presence of three emission features; two of which
correspond to both components of the
system with the stronger coming from the hotter star.
 The third emission component is
blue-shifted by 3.4~\AA$\ $ with respect to the hotter
emission component and  could be related
with the mass transfer from the cool component to the hot one.

We have observed this system in the Ca~{\sc ii} H \& K line region in two
epochs. One spectrum taken in Feb-88 at orbital phase 0.36 (FFMCC) and a new
observation taken in Mar-93 at orbital phase 0.44.
Both spectra show strong emission which is normally attributed to the hotter
component, since this one has the main contribution to the observed spectrum.
However the spectral subtraction technique reveals that the cool component
also contributes to this emission (in the spectrum at phase 0.36,
Fig.~\ref{fig:hykrzcnc}, upper panel)
 and that the hot and more active star shows a small H$\epsilon$ emission line.
In the spectrum at phase 0.44 (Fig.~\ref{fig:hykrzcnc}, lower panel)
the difference in radial velocity between the
components is very small and it is not possible to separate both
contributions.


\subsubsection{DM UMa (BD +61 1211)}


DM UMa is a single-lined spectroscopic binary and
one of the few RS CVn binaries which show H$\alpha$
emission above the continuum at all times
(Mohin \& Raveendran 1992, 1994; Hatzes 1995).

Our spectra in the H$\alpha$ line region (Montes et al. 1995a, b)
show a strong and asymmetric H$\alpha$ emission above the continuum.

One spectrum in Ca~{\sc ii} H \& K line region, taken in Mar-93 at
0.85 orbital phase (Fig.~\ref{fig:hykdmuma}, upper panel), shows a strong
emission and the presence of the H$\epsilon$ line also in emission.
The synthesized spectrum has been constructed with a K1IV reference star.
The subtracted spectrum reveals that H$\zeta$
and H$\eta$ Balmer lines appear also in emission
(Fig.~\ref{fig:hykdmuma}, lower panel), which indicate that DM UMa
is a very active star in agreement with the H$\alpha$ observations.

In the subtracted spectrum we can see that the excess Ca~{\sc ii} H \& K
emission line profiles present pronounced wings and
are not well matches with Gaussian profiles. This result indicates the
presence of a broad component in the excess Ca~{\sc ii} H \& K emission
lines similar to that
found by us and Hatzes (1995) in the excess H$\alpha$  emission.


\subsubsection{DK Dra (HD 106677, HR 4665)}


This is a double-lined spectroscopic binary with almost identical
components of spectral type K1III
 and  Ca~{\sc ii} H \& K emissions from both components
 (Bopp et al. 1979; Fekel et al. 1986; Strassmeier 1994b).
Eker et al. (1995) reported a variable nature of H$\alpha$ and,
using a subtracted spectrum, found emission of similar
intensity from both components.

{}From one spectrum taken in the H$\alpha$ line region (Montes et al. 1995a, b)
we found  a broad excess emission feature as a
result of the emission from both components.

Our observations in the  Ca~{\sc ii} H \& K line region cover three observing
seasons: one observation taken in Nov-86 at orbital phase 0.44 (FFMCC), three
observation in Jan-88 at orbital phases 0.10 and 0.13 (FFMCC), and two new
observations taken in Mar-93 at phases 0.02 and 0.05.
Because our orbital phases are very near to 0.0 and 0.5, it is not possible to
separate the contribution from both components.
In Fig.~\ref{fig:hykdkdra} we can see that the emission
line profile show small variations  with
the phase and, more important, time variations from Nov-86 to Mar-93.

The spectral subtraction technique points out that the spectra of Mar-93,
which present the higher levels of activity, also show the H$\epsilon$ line in
emission.


\subsubsection{GX Lib (HD 136905) }


Single-lined spectroscopic binary of spectral type K1III, previously studied by
us in the Ca~{\sc ii} H \& K and H$\alpha$ lines (FFMCC; Montes et al. 1995a).

Two spectra taken in Jul-89 at orbital phases 0.44 and 0.36, (FFMCC)
 show a wide emission line at the center of the absorption line.
A new observation of this system taken in Mar-93 at orbital phase 0.83
(Fig.~\ref{fig:hykgxlib}, upper panel) confirms this
behaviour. Small variations in the emission fluxes and in the line widths
 in these three spectra were observed (Fig.~\ref{fig:hykgxlib}).
The synthesized spectrum has been constructed with a K1IV reference star.


\subsubsection{4 UMi (HD 124547, HR 5321)}


Single-lined spectroscopic binary of spectral type K3III and with a long
orbital period (605.08 days).
This system is listed in the first edition of CABS
(Strassmeier et al. 1988) as having
weak Ca~{\sc  ii} H \& K emission lines (I$_{\rm K}$=3, Wilson 1976).
However it is not included in the
second edition of CABS due to insufficient evidence for
chromospheric activity in  the  Ca~{\sc  ii} H \& K lines
(Strassmeier et  al. 1990) and in the H$\alpha$ line
(Frasca \&  Catalano 1994).
However, Xuefu et al. (1993) considered that this system is
a chromospherically active star, because the H$\alpha$ line is filled by
emission.

In our spectrum in the  Ca~{\sc  ii} H \& K line
region taken in Mar-93 at orbital phase 0.86
(Fig.~\ref{fig:hyk4umi}) we have found
a weak emission in these lines.
The application of the spectral subtraction
technique also pointed out the presence of this weak emission in the H \& K
lines.


\subsubsection{DR Dra (29 Dra, HD 160538)}


Single-lined spectroscopic binary consisting of a hot white dwarf
and a K0-2III primary (Fekel \& Simon 1985).
The K star is a highly asynchronous rotator and has been
previously studied by us in the Ca~{\sc ii} H \& K
and H$\alpha$ lines (FFMCC; Montes et al. 1995a).

Two observations in the  Ca~{\sc ii} H \& K line region
 of this systems are available. One spectrum taken in Jul-89 at orbital
phase 0.67 (FFMCC) and a new spectrum taken in Mar-93 at phase 0.09
(Fig.~\ref{fig:hykdrdra}).
Both spectra show strong H \& K emissions above the continuum without
considerable variation of the emission line fluxes between these two epochs.
By subtracting the synthesized spectrum we found a small
excess H$\epsilon$ emission.


\subsection{Single chromospherically active stars}

\subsubsection{V2213 Oph (HD 154417, HR 6349)}

This is a single star of F8.5IV-V spectral type
and with a rotational period of 7.6 days,
obtained from modulation of the S index (Noyes et al. 1984).

We have taken one spectrum of this system in Jul-89 which
shows weak emission in the Ca~{\sc ii} H \& K lines with an intensity
I$_{\rm K_{3}}$=0.21 (Fig.~\ref{fig:hykv2213oph}).
The subtraction of a F8V reference star confirms
this small emission.

\subsubsection{59 Vir (HD 115383, HR 5011)}

59 Vir is a single active G0V star with a 3.33 day rotational period
(Noyes et al. 1984) and was also observed in the H$\alpha$
line by Herbig (1985).

Our spectrum in the Ca~{\sc ii} H \& K line region (Fig.~\ref{fig:hyk59vir}),
taken in Jul-89,
shows a weak emission line (I$_{\rm K_{3}}$=0.31) similar to that observed
by Linsky et al. (1979).
The synthesized spectrum has been constructed with a G0V reference star.

\subsubsection{HN Peg (HD 206860, HR 8314)}

Another G0V single star
with a rotational period of 4.7 days (Noyes  et  al.  1984).

In our spectrum in the  Ca~{\sc  ii} H \& K line
region, taken in Jul-89, we have found
a weak but clear emission in these lines (Fig.~\ref{fig:hykhnpeg})
with an intensity I$_{\rm K_{3}}$=0.35.
The spectrum is well matched using a reference  star  of  spectral
type G0V.

\subsubsection{$\xi$ UMa A (53 UMa, HD 98231, HR 4375)}

This star is the A-component of the visual binary ADS 8119  AB, whose
B-component is a RS CVn system already studied by us (FFMCC).
$\xi$ UMa~A, is a SB with an orbital period
 of 669.17 days and a G0V spectral type.
Wilson (1963) noted the Ca~{\sc ii} H \& K lines  in $\xi$ UMa~B,
while the A component shows no detectable emission.
However, Wooley et al. (1970) found a weak emission in this component.

Our spectrum of this star, in the Ca~{\sc ii} H \& K line region,
taken in Jul-89, shows a very weak emission (Fig.~\ref{fig:hykxiumaa}),
 which is possible to measure using the
reconstruction of the absorption line profile and the spectral subtraction
technique.

\subsubsection{$\sigma$$^{1}$ CrB (HD 146362, HR 6364)}

$\sigma$$^{1}$ CrB is the fainter member of the visual binary ADS 9979.
The brighter visual companion $\sigma$$^{2}$ CrB is a
chromospherically active binary previously studied by us (FFMCC).

Our spectrum of this G1V star, in the Ca~{\sc ii} H \& K line region,
 taken in Feb-88, seems not to present any
indication of chromospheric activity (Fig.~\ref{fig:hyks1crb}),
 however, by subtraction of the
synthesized spectrum, constructed with a G1V reference star
we have found a weak emission in these lines.

\subsubsection{$\kappa$$^{1}$ Cet  (HD 20630, HR 996)}

$\kappa$$^{1}$ Cet is a single active G5V star with a 9.4 day rotational period
and extensively studied in the Ca~{\sc ii} H \& K lines in the literature
(Pasquini et al. (1988), Pasquini (1992)
y Garc\'{\i}a L\'{o}pez et al. (1990, 1992).

Our spectrum in the Ca~{\sc ii} H \& K line region (Fig.~\ref{fig:hykk1cet}),
taken in Dec-92, shows a moderate emission in these lines
(I$_{\rm K_{3}}$=0.40).
The synthesized spectrum has been constructed with a G1V reference star.

\subsubsection{$\xi$ Boo A and B  (HD 131156 A and B, HR 5544 A and B)}

Visual binary (ADS 9413) composed of stars of spectral types
G8V (component A) and K4V (component B)
and rotational periods of 6.2 and 11.5 days, respectively
(Noyes et al. 1984).
$\xi$~Boo~B has the largest amplitude of chromospheric activity in the
S index (2.1 to 1.2) (Baliunas et al. 1995)

We have taken individual spectra in the Ca~{\sc ii} H \& K line
region of both components of this visual binary.
The H \& K emission lines of both components are very different.
While $\xi$~Boo~A shows weak emission lines (I$_{\rm K_{3}}$=0.63),
$\xi$~Boo~B exhibits very strong emissions well above the nearby stellar
continuum (I$_{\rm K_{3}}$=2.51) and also presents the H$\epsilon$ line in
emission.
The spectrum of $\xi$~Boo~A is well matched using a reference  star  of
spectral type G8IV (Fig.~\ref{fig:hykxibooab}, upper panel).
 However for the spectrum of $\xi$~Boo~B
(Fig.~\ref{fig:hykxibooab}, lower panel)
we have not found a satisfactory fit with the synthesized spectra constructed
with a K3V reference star.

\subsubsection{61 UMa  (HD 101501, HR 4496)}

Single active star of spectral type G8V, which is included
in the list of standards of Taylor (1984).
It presents modulation of the Ca~{\sc ii} S index and a
rotational period of 17.1 days (Noyes et al. 1984).

Seven spectra of this star in the Ca~{\sc ii} H \& K line
region, taken in Mar-93, are available.
In these spectra (Fig.~\ref{fig:hyk61uma}) we observe a
weak emission  (I$_{\rm K_{3}}$=0.30)
very similar to that observed by Linsky et al. (1979) and
Strassmeier et al. (1990).
The synthesized spectrum has been constructed with a G8V reference star.

\subsubsection{$\epsilon$ Eri  (HD 22049, HR 1084)}

$\epsilon$ Eri is a single active star classified as K2V  with a rotational
period of 11.3 days (Noyes et al. 1984).
The  Ca~{\sc ii} H \& K lines have been studied by Linsky et al. (1979),
 Zarro \& Rodgers (1983) and Garc\'{\i}a L\'{o}pez et al. (1992).

Our spectrum in the Ca~{\sc ii} H \& K region, taken in Dec-92, shows a
moderate emission (I$_{\rm K_{3}}$=0.80).
The spectrum is well matched using a reference  star  of  spectral
type K1V (Fig.~\ref{fig:hykeeri}).

\subsubsection{HD 4628 (HR 222)}

Single dwarf star classified as K4V by Noyes et al. (1984) and listed as K2V
in the Bright Star Catalogue (Hoffleit \& Jaschek 1982).
Recently, Mathioudakis et al. (1994) reported the detection
of EUV emission from this
star and suggested the existence of a cool corona.

We have taken one spectrum of HD 4628 in Dec-92 which shows weak emission
in the Ca~{\sc ii} H \& K lines (I$_{\rm K_{3}}$=0.29).
We have used a K1V reference star to perform the spectral
subtraction (Fig.~\ref{fig:hykhd4628}).

\subsubsection{HD 115404 }

Another K dwarf star with a rotational period of 18.3 days
(Noyes et al. 1984) and studied in the Ca~{\sc ii} H \& K region by
Thatcher \& Robinson (1993).

Our spectrum in the Ca~{\sc ii} H \& K line region, taken in Jul-89, shows
moderate emission (I$_{\rm K_{3}}$=0.80) (Fig.~\ref{fig:hykhd115404}).
The synthesized spectrum has been constructed with a K1V reference star.

\subsubsection{$\rho$ Boo (25 Boo, HD 127665, HR 5429)}

K3 giant included in the catalog of Keenan \& McNeil (1989).
Observed in the Ca~{\sc  ii} H \& K line region by
Strassmeier et al. (1990) as a non-active chromosphere star.

We have taken one spectrum of this star in Mar-93 in the
Ca~{\sc  ii} H \& K line region (Fig.~\ref{fig:hykrboo}).
This spectrum shows very weak emissions in these lines
(I$_{\rm K_{3}}$=0.10). This star has not been used as reference star.

\subsubsection{61 Cyg A and B (HD 201091 and 201092, HR 8085 and 8086)}

This visual binary has components of spectral type K5V and K7V and rotational
periods of 37.9 and 48.0 days, respectively (Noyes et al. 1984).
Both components present Ca~{\sc ii} H \& K emission lines (Linsky et al. 1979;
 Strassmeier et  al. 1990).

In Jul-89 we have taken spectra in the Ca~{\sc ii} H \& K line
region of both components of this visual binary.
The H \& K emission lines of both components are different,
61 Cyg A presents an emission line with an intensity I$_{\rm K_{3}}$=1.1,
while 61 Cyg B exhibits very strong emissions well above the nearby stellar
continuum (I$_{\rm K_{3}}$=2.0).
Both components show also the H$\epsilon$ line in emission, but it is more
dificult to see in the B component.


\begin{acknowledgements}

This work has been supported by the Universidad Complutense de Madrid
and the Spanish Direcci\'{o}n General de Investigaci\'{o}n
Cient\'{\i}fica y  T\'{e}cnica (DGICYT) under grant PB91-0348.
We would like to thank the referee K.G. Strassmeier
for suggesting several improvements and clarifications.

\end{acknowledgements}




\begin{table*}[p]
\caption[]{Stellar parameters, chromospheric active binaries
(Groups 1, 2 and 3)}
\begin{flushleft}
\scriptsize
\begin{tabular}{l l c c c c c l l c }
\noalign{\smallskip}
\hline
\noalign{\smallskip}
{HD} & {Name} & {T$_{\rm sp}$} & {SB} & {R} &
 {d} & {V-R} & {P$_{\rm orb}$} & {P$_{\rm rot}$} & Vsin{\it i}\\
             &    &     &    & (R$_\odot$) & (pc) &   & (days) &
 (days) & (km s$^{-1}$)  \\
\noalign{\smallskip}
\hline
Group 1. \\
\hline
\noalign{\smallskip}
%
 17433 & VY Ari  &  K3-4V-IV     & 1  & -        & 21 & 0.61 & 13.198 & 16.42
& 6 \\
45088  &  OU Gem   &  K3V/K5V    & 2  & -       & 12 & 0.82/0.99 & 6.9919
& 7.3600 & 5.6/5.6 \\
-     & YY Gem   &  dM1e/dM1e    & 2  & 0.62/   & 13.7 & 1.40/1.40 & 0.8142822
& 0.8143 & 40/40 \\
80715 & BF Lyn   &  K2V/[dK]     & 2  & $\geq$0.78/$\geq$0.78 & 29 &
& 3.80406  & $\approx$P$_{\rm orb}$ & 10/10 \\
86590 & DH Leo   & $\{$K0V/K7V$\}$/K5V&  3 & $\{$0.97/0.67$\}$ & 32
& 0.64/0.90  &  1.070354 & 1.0665 & $\{$45/31$\}$/8 \\
107760 & AS Dra   &  G4V/G9V    & 2 &  -       & 29.4 & 0.60  &  5.414905
& $\approx$P$_{\rm orb}$ & 12/8  \\
108102  & IL Com   &  F8V/F8V    & 2 & [1.1/1.1] & 86 & 0.47/0.47 & 0.9620
& 0.8200 & 35/35 \\
131511  & HR 5553  &  K2V        & 1 &  -        & 11.9  &  0.74
& 125.369 &   -     &  4    \\
143313  & MS Ser   &  K2V/K6V    & 2 &  -        & 30    &
& 9.0149  & 9.60 &  -      \\
218738  &  KZ And   & dK2/dK2    &  2 & $\geq$0.74/ &[$\approx$23]
&  [0.74/0.74] & 3.032867 & 3.03 & 12.3/11.6   \\
222317  & KT Peg    & G5V/K6V    &  2 & 0.93/0.72  & 25 &
& 6.20199 & 6.092  & 8/5 \\
%
\noalign{\smallskip}
\hline
Group 2. \\
\hline
\noalign{\smallskip}
%
 21242 & UX Ari  &  G5V/K0IV     & 2  & 0.93/$\geq$4.7 & 50 &  0.70/0.54
& 6.43791 & $\approx$P$_{\rm orb}$ & 6/37 \\
106225  & HU Vir   & K0IV        & 1  & $\geq$5.7  & 220 &
&  10.3876 & 10.28  & 25   \\
113816  & BD-04 3419 &  K2IV-III & 1 & -       & 165   &         &   $>$ 20
& -      &  6     \\
\noalign{\smallskip}
\hline
Group 3. \\
\hline
\noalign{\smallskip}
%
352    & 5 Cet   &  F/K1III      & 1  & $\approx$41 & 140 &  0.81   & 96.439
& 48.16$^{a}$ & /22 \\
1833   & BD Cet  &  K1III        & 1  & $\geq$10       & 71  &  0.81
& 35.1   & 34.46 & 15 \\
4502   & $\zeta$ And & /K1III     & 1  & $\approx$0.7/13.4 & 31 &  0.84
& 17.7692 & 8.917$^{a}$ & 41  \\
5516   & $\eta$ And &G8IV-III/G8IV-III & 2 &  -     & 111 &  0.94 & 115.71
& -  & 15   \\
7672   & AY Cet  &   WD/G5III    & 1  & 0.012/15  & 66.7  &         &  56.824
& 77.22 & 4 \\
12545  & BD+34 363 & K0III       & 1  & $\geq$8  & 310 &         &  23.9824
& 24.3 & 17  \\
13480  & 6 Tri    & F5/K0III    & 2  & 13       & 75  & [/0.77] & 14.7339
& $\approx$P$_{\rm orb}$ & 7/34 \\
37824  & V1149 Ori &  K1III      & 1  & $\geq$11     & [164] & 0.90  & 53.58
& 54.1 & 11 \\
73343 &  RZ Cnc  &  K1III/K3-4III & 2 & 10.2/12.2 & 395  & 0.81/0.96 & 21.6430
& $\approx$P$_{\rm orb}$ & 25/22  \\
-     &  DM UMa  &  K0-1IV-III   & 1 &  $\geq$3.8 &  130 & 0.80 & 7.4949
&  7.478 & 36 \\
 106677 & DK Dra   & K1III/K1III & 2 & $\geq$13/$\geq$13 & 130 & 0.89/0.89
& 64.44 &  63.75 & 10/10 \\
124547  & 4 UMi    & K3III       & 1 &  -      & 100&  0.96   & 605.8
& 160.0  &  15  \\
136905  & GX Lib   & [G-KV]/K1III& 1 & /$\geq$7  & [219] &     0.84
& 11.1345 & 11.134  &  /32    \\
160538  & DR Dra   & WD/K0-2III  & 1 & 0.012/$\geq$5 & 87.9  & [/0.81] & 39.
& 31.5 &  /8    \\
%
\noalign{\smallskip}
\hline
\noalign{\smallskip}
\end{tabular}
\end{flushleft}
\end{table*}

\begin{table*}[p]
\caption[]{Stellar parameters of reference stars, active
single stars and active components of visual binaries.
\label{tab:parwilson} }
\begin{flushleft}
\scriptsize
\begin{tabular}{l l l c c c c}
\hline
\noalign{\smallskip}
 HD & Name &  T$_{\rm sp}$   & V-R & P$_{\rm rot}$ & Vsin{\it i} & A/R$^{*}$ \\
    &        &             &     & (days)   & (km s$^{-1}$) & \\
\hline
 {\bf F }  \\
\hline
\noalign{\smallskip}
 13480B  & ADS 1697 B &   F5V   & 0.40 & 2.236& -    & R \\
 120136& $\tau$ Boo&   F6IV  & 0.42 & -    & 10.0 & R \\
 124850  &$\iota$ Vir& F6III &      &      & 15.0 & R \\
 187013  & 17 Cyg  &   F7V   & 0.45 & -    & 10.0 & R \\
 212754  & 34 Peg  &   F7V   & 0.45 & -    & 10.0 & R \\
 216385 & $\sigma$ Peg&F7IV  & 0.45 & -    & 10.0 & R \\
 187691 & o Aql  &     F8V   & 0.47 & -    & 5.0  & R \\
 142373 & $\chi$ Her&   F8V  & 0.47 & -    & 10.0 & R \\
 194012 & HR 7793&      F8V  & 0.47 & -    & 5.0  & R \\
 45067  & HR 2313&      F8V  & 0.47 & -    & $<$ 15 & R \\
 6920   & 44 And &      F8V  & 0.47 & 15.3 & $<$ 15 & R \\
 107213 & 9 Com  &      F8V  & 0.47 & -    & 10.0 & R \\
 136202 & 5 Ser  &  F8III-IV & 0.48 & -    & 5.0  & R \\
 154417 & V2213 Oph&  F8.5IV-V & 0.48 & 7.78 & 5.0 & A  \\
 43587  & HR 2251&      F9V  & 0.49 & -    & 5.0  & R \\
\noalign{\smallskip}
\hline
{\bf G} \\
\hline
\noalign{\smallskip}
115383  & 59 Vir  & G0V   & 0.50& 3.33  & 5.0  & A \\
152792  &      & G0V   &    0.50& -     & -    & R \\
114710  & $\beta$ Com & G0V&0.50& 12.35 & 10.0 & R \\
206860  & HN Peg & G0V   &  0.50& 4.86  & 10.0 & A \\
29645   & HR 1489& G0V   &  0.50& -     & $<$ 15 & R \\
13974   & $\delta$ Tri& G0V&0.50& -     & 10.0 & R \\
98231   & $\xi$ UMa A&G0V & 0.50& -     & $<$ 15 & A \\
218739  & ADS 16557 A&G0V&  0.50& -     & -         & A \\
13421   & 64 Cet & G0IV  &  0.50& -     & $<$  15& R \\
190406  & 15 Sge & G1V   &  0.52& 13.94 & 5.0    & R \\
146362 & $\sigma$$^{1}$ CrB & G1V & 0.52& - & -  & R \\
-       & Sun    & G2V   &  0.53& 25.72 &        & R \\
143761  & $\rho$ CrB & G2V &0.53& 0.0   & 5.0    & R \\
81809   & HR 3750& G2V   &  0.53& 40.20 & 10.0   & R \\
9562    & HR 448 & G2IV  &  0.61& -     &  $<$  15&R \\
12235   & 112 Psc& G2IV  &  0.61& -     &  $<$  15&R \\
217014  & 51 Peg  & G2.5IV& 0.61& -     & 0       &R \\
20630&$\kappa^{1}$ Cet& G5V&0.54& 9.24  & $<$  15 &A \\
115617  & 61 Vir & G6V   &  0.55& -     & $<$  15 &R \\
190360  & HR 7670& G6IV  &  0.62& -     & -       &R \\
182488  & HR 7368& G8V   &  0.58& -     & -       &R \\
131156A & $\xi$ Boo A& G8V&0.58& 6.31  & 3        &A \\
144287  &      & G8V   &    0.58& -     & -       &R \\
101501  & 61 UMa & G8V   &  0.58& 16.68 & $<$  15 &A \\
182572  & 31 Aql & G8IV   & 0.64& -     & $<$  15 &R \\
188512  & $\beta$ Aql& G8IV&0.64& -     & 15      &R \\
158614  & HR 6516& G8IV   & 0.64& -     & -       &R \\
\noalign{\smallskip}
\hline
 {\bf K} \\
\hline
\noalign{\smallskip}
3651    & 54 Psc & K0V  &   0.64& 48.00 & -       & R \\
50082 &BD+06 1411& K0III &  0.77& -     & -       & R \\
190404  &        & K1V   &  0.69& -     & -       & A \\
10476   & 107 Psc& K1V    & 0.69& 35.2  & $<$  20 & R \\
22072   & HR 1085& K1IV  &  0.75& -     & -       & R \\
142091  &$\kappa$ CrB& K1IV& 0.75& -    & $<$  15 & R \\
50843 &BD+04 1506 &K1III   & 0.81& -    &         & R \\
22049 &$\epsilon$ Eri& K2V &0.74& 11.68 & $<$  15 & A \\
4628    & HR 222 & K2V   &  0.74& 38.5  & -       & A \\
16160   & HR 753 & K3V   &  0.82& 48.0  & -       & A \\
219134  & HR 8832& K3V   &  0.82& -     & -       & A \\
115404  &        & K3V   &  0.82& 18.47 & -       & A \\
127665  &$\rho$ Boo&K3III & 0.96& -     & $<$  15 & A \\
131156B & $\xi$ Boo B& K4V&0.91& 12.28 & 20       & A \\
201091 & 61 Cyg A  &  K5V & 0.99& 35.37 & 10      & A \\
201092 & 61 Cyg B  &  K7V & 1.15& 37.84 & $<$  25 & A \\
\noalign{\smallskip}
\hline
\noalign{\smallskip}
\end{tabular}

\vspace{0.4cm}

$^{*}$ A: Active star / R: Reference star

\end{flushleft}
\end{table*}

\begin{table*}[p]
\caption[]{Ca~{\sc ii} H \& K lines measures in the observed and
subtracted spectrum  (Group 1)
\label{tab:medhyk1} }
\begin{flushleft}
\scriptsize
\begin{tabular}{l c c c c c c c c c c c c c c c}
\hline
\noalign{\smallskip}
 &    &  &   &  & \multicolumn{4}{c}{Reconstruction of the profile} &\ &
\multicolumn{6}{c}{Spectral subtraction} \\
\cline{6-9}\cline{11-16}
\noalign{\smallskip}
\noalign{\smallskip}
 Name  & Date & $\varphi$ & E & S$_{\rm H}$/S$_{\rm C}$ &
EW & C(K) & EW & C(H) & &
EW & I(K) & EW & I(H) & EW & I(H$\epsilon$) \\
 & & & & & (K) & & (H) & & & (K) &  & (H) &   & (H$\epsilon$) &  \\
\noalign{\smallskip}
\hline
\noalign{\smallskip}
          &          &      &   &           &       &      &       &      &
&       &      &       &      &       &      \\
 VY Ari   & 16/12/92 & 0.17 & - &     -     & 1.643 & 3.22 & 1.412 & 2.51 &
& 1.789 & 2.46 & 1.614 & 2.27 & 0.477 & 0.36 \\
          &          &      &   &           &       &      &       &      &
&       &      &       &      &       &      \\
 OU Gem   & 04/03/93 & 0.47 & T & 0.74/0.26 & 0.774 & 1.78 & 0.715 & 1.50 &
& 0.938 & 1.09 & 0.920 & 1.02 & 0.232 & 0.26 \\
          &          &      &   &           &       &      &       &      &
&       &      &       &      &       &      \\
 YY Gem   & 04/03/93 & 0.44 & 1 &    0.50   & 5.049 & 3.25 & 4.901 & 1.91 &
& 5.232 & 5.12 & 5.378 & 4.82 & -     & -    \\
          &          &      & 2 &    0.50   & 6.033 & 3.74 & 7.613 & 2.61 &
& 6.209 & 6.32 & 7.104 & 7.11 & 3.199 & 2.04 \\
          &          &      &   &           &       &      &       &      &
&       &      &       &      &       &      \\
 BF Lyn   & 05/03/93 & 0.21 & H &    0.50   & 0.832 & 1.80 & 0.717 & 1.34 &
& 0.891 & 1.25 & 0.861 & 1.16 & -     & -    \\
          &          &      & C &    0.50   & 0.823 & 1.86 & 0.795 & 1.11 &
& 0.881 & 1.19 & 0.982 & 1.27 & 0.376 & 0.24 \\
          &          &      &   &           &       &      &       &      &
&       &      &       &      &       &      \\
 DH Leo   & 29/01/88 & 0.55 & H &    0.94   & 0.864 & 1.14 & 0.788 & 0.92 &
& 1.075 & 0.93 & 1.111 & 0.94 & 0.268 & 0.18 \\
          &          &      & C &    0.06   & 0.203 & 0.34 & 0.210 & 0.29 &
& 0.218 & 0.22 & 0.183 & 0.24 & -     & -    \\
 "        & 01/02/88 & 0.32 & H &    0.94   & 1.469 & 1.32 & 1.412 & 0.90 &
& 1.017 & 0.93 & 1.008 & 0.94 & 0.245 & 0.21 \\
          &          &      & C &    0.06   & 0.446 & 0.40 & 0.160 & 0.22 &
& 0.243 & 0.30 & 0.276 & 0.30 & -     & -    \\
 "        & 05/03/93 & 0.07 & H &    0.94   & 0.948 & 1.13 & 0.804 & 0.89 &
& 1.098 & 0.94 & 0.965 & 0.85 & 0.314 & 0.21 \\
          &          &      & C &    0.06   & 0.164 & 0.22 & 0.325 & 0.33 &
& 0.256 & 0.28 & 0.214 & 0.23 & -     & -    \\
 "        & 08/03/93 & 0.70 & H &    0.94   & 1.067 & 0.97 & 0.823 & 0.49 &
& 1.085 & 0.88 & 1.133 & 0.79 & 0.213 & 0.15 \\
          &          &      & C &    0.06   & 0.283 & 0.31 & 0.297 & 0.23 &
& 0.363 & 0.25 & 0.406 & 0.38 & -     & -    \\
 "        & 07/03/93 & 0.87 & H &    0.94   & 1.052 & 1.18 & 0.753 & 0.57 &
& 1.038 & 0.83 & 1.074 & 0.85 & 0.240 & 0.19 \\
          &          &      & C &    0.06   & 0.355 & 0.46 & 0.068 & 0.10 &
& 0.256 & 0.24 & 0.253 & 0.21 & -     & -    \\
          &          &      &   &           &       &      &       &      &
&       &      &       &      &       &      \\
 AS Dra   & 07/03/93 & 0.49 & T & 0.66/0.34 & 0.311 & 0.84 & 0.281 & 0.79 &
& 0.444 & 0.54 & 0.383 & 0.50 & 0.208 & 0.12 \\
 "        & 09/03/93 & 0.85 & H &    0.66   & 0.174 & 0.48 & 0.205 & 0.53 &
& 0.267 & 0.30 & 0.272 & 0.31 & -     & -    \\
          &          &      & C &    0.34   & 0.152 & 0.41 & 0.137 & 0.31 &
& 0.234 & 0.24 & 0.235 & 0.23 & -     & -    \\
          &          &      &   &           &       &      &       &      &
&       &      &       &      &       &      \\
 IL Com   & 31/01/88 &   *  & 1 &    0.50   & 0.143 & 0.21 & 0.116 & 0.15 &
& 0.315 & 0.16 & 0.330 & 0.17 & -     & -    \\
          &          &      & 2 &    0.50   & 0.108 & 0.18 & 0.069 & 0.11 &
& 0.177 & 0.11 & 0.184 & 0.12 & -     & -    \\
 "        & 01/02/88 &   *  & 1 &    0.50   & 0.136 & 0.24 & 0.087 & 0.14 &
& 0.302 & 0.18 & 0.349 & 0.21 & -     & -    \\
          &          &      & 2 &    0.50   & 0.111 & 0.19 & 0.056 & 0.22 &
& 0.181 & 0.14 & 0.180 & 0.14 & -     & -    \\
 "        & 05/03/93 &   *  & 1 &    0.50   & 0.115 & 0.19 & 0.072 & 0.11 &
& 0.183 & 0.14 & 0.154 & 0.12 & -     & -    \\
          &          &      & 2 &    0.50   & 0.144 & 0.20 & 0.095 & 0.14 &
& 0.250 & 0.17 & 0.297 & 0.16 & -     & -    \\
          &          &      &   &           &       &      &       &      &
&       &      &       &      &       &      \\
 HD 131511& 05/03/93 & 0.75 & - &     -     & 0.259 & 0.82 & 0.200 & 0.58 &
& 0.348 & 0.39 & 0.317 & 0.35 & -     & -    \\
          &          &      &   &           &       &      &       &      &
&       &      &       &      &       &      \\
 MS Ser   & 07/03/93 & 0.16 & H & 0.82/0.18 & 1.965 & 3.00 & 1.719 & 2.24 &
& 2.004 & 2.13 & 1.832 & 2.03 & 0.535 & 0.33 \\
          &          &      &   &           &       &      &       &      &
&       &      &       &      &       &      \\
 KZ And   & 07/12/89 & 0.33 & H &    0.50   & 0.600 & 1.34 & 0.544 & 1.15 &
& 0.757 & 1.17 & 0.709 & 1.09 & -     & -    \\
          &          &      & C &    0.50   & 0.568 & 1.31 & 0.609 & 1.08 &
& 0.738 & 1.16 & 0.845 & 1.25 & 0.329 & 0.30 \\
 "        & 15/12/92 & 0.39 & H &    0.50   & 0.631 & 1.56 & 0.625 & 1.49 &
& 0.734 & 1.07 & 0.631 & 0.92 & -     & -    \\
          &          &      & C &    0.50   & 0.605 & 1.41 & 0.574 & 1.07 &
& 0.716 & 1.03 & 0.695 & 0.97 & 0.308 & 0.23 \\
          &          &      &   &           &       &      &       &      &
&       &      &       &      &       &      \\
 KT Peg   & 15/12/92 & 0.27 & H &    0.90   & 0.184 & 0.62 & 0.141 & 0.44 &
& 0.243 & 0.30 & 0.192 & 0.24 & -     & -    \\
          &          &      & C &    0.10   & 0.095 & 0.36 & 0.084 & 0.33 &
& 0.088 & 0.09 & 0.169 & 0.08 & -     & -    \\
          &          &      &   &           &       &      &       &      &
&       &      &       &      &       &      \\
\noalign{\smallskip}
\hline
\end{tabular}
\end{flushleft}
\end{table*}
\begin{table*}[p]
\caption[]{Ca~{\sc ii} H \& K lines measures in the observed and
subtracted spectrum  (Group 2)
\label{tab:medhyk2} }
\begin{flushleft}
\scriptsize
\begin{tabular}{l c c c c c c c c c c c c c c c}
\hline
\noalign{\smallskip}
 &    &  &   &  & \multicolumn{4}{c}{Reconstruction of the profile} &\ &
\multicolumn{6}{c}{Spectral subtraction} \\
\cline{6-9}\cline{11-16}
\noalign{\smallskip}
\noalign{\smallskip}
 Name  & Date & $\varphi$ & E & S$_{\rm H}$/S$_{\rm C}$ &
EW & C(K) & EW & C(H) & &
EW & I(K) & EW & I(H) & EW & I(H$\epsilon$) \\
 & & & & & (K) & & (H) & & & (K) &  & (H) &   & (H$\epsilon$) &  \\
\noalign{\smallskip}
\hline
\noalign{\smallskip}
          &          &      &   &           &       &      &       &      &
&       &      &       &      &       &      \\
 UX Ari   & 16/12/93 & 0.92 & C & 0.60/0.40 & 1.522 & 1.36 & 1.332 & 1.87 &
& 1.609 & 1.29 & 1.475 & 1.21 & 0.292 & 0.24 \\
          &          &      &   &           &       &      &       &      &
&       &      &       &      &       &      \\
 HU Vir   & 09/03/93 & 0.71 & - &     -     & 2.375 & 3.45 & 2.608 & 4.07 &
& 2.674 & 2.39 & 2.573 & 2.46 & 0.794 & 0.47 \\
          &          &      &   &           &       &      &       &      &
&       &      &       &      &       &      \\
 HD 113816& 07/03/93 & 0.68 & - &     -     & 2.815 & 6.01 & 2.461 & 3.04 &
& 2.892 & 2.49 & 2.650 & 2.02 & -     & -    \\
          &          &      &   &           &       &      &       &      &
&       &      &       &      &       &      \\
\noalign{\smallskip}
\hline
\end{tabular}
\end{flushleft}
\end{table*}
\begin{table*}[p]
\caption[]{Ca~{\sc ii} H \& K lines measures in the observed and
subtracted spectrum  (Group 3)
\label{tab:medhyk3} }
\begin{flushleft}
\scriptsize
\begin{tabular}{l c c c c c c c c c c c c c c c}
\hline
\noalign{\smallskip}
 &    &  &   &  & \multicolumn{4}{c}{Reconstruction of the profile} &\ &
\multicolumn{6}{c}{Spectral subtraction} \\
\cline{6-9}\cline{11-16}
\noalign{\smallskip}
\noalign{\smallskip}
 Name  & Date & $\varphi$ & E & S$_{\rm H}$/S$_{\rm C}$ &
EW & C(K) & EW & C(H) & &
EW & I(K) & EW & I(H) & EW & I(H$\epsilon$) \\
 & & & & & (K) & & (H) & & & (K) &  & (H) &   & (H$\epsilon$) &  \\
\noalign{\smallskip}
\hline
\noalign{\smallskip}
          &          &      &   &           &       &      &       &      &
&       &      &       &      &       &      \\
 5 Cet    & 15/12/92 & 0.32 & C &     -     & 0.440 & 0.67 & 0.273 & 0.48 &
&  ...  & ...  &  ...  & ...  & -     & -    \\
          &          &      &   &           &       &      &       &      &
&       &      &       &      &       &      \\
 BD Cet   & 12/12/92 & 0.90 & C &     -     & 0.984 & 1.89 & 0.699 & 1.16 &
& 1.147 & 0.88 & 1.005 & 0.82 & -     & -    \\
          &          &      &   &           &       &      &       &      &
&       &      &       &      &       &      \\
$\zeta$ And&24/10/91 & 0.29 & - &     -     & 0.715 & 0.98 & 0.568 & 0.74 &
& 0.945 & 0.58 & 0.916 & 0.56 & -     & -    \\
 "        & 12/12/92 & 0.69 & - &     -     & 0.784 & 1.39 & 0.636 & 0.87 &
& 0.980 & 0.66 & 0.957 & 0.61 & -     & -    \\
          &          &      &   &           &       &      &       &      &
&       &      &       &      &       &      \\
$\eta$ And& 12/12/92 & 0.62 & T & 0.50/0.50 & 0.102 & 0.51 & 0.065 & 0.31 &
& 0.077 & 0.08 & 0.057 & 0.08 & -     & -    \\
          &          &      &   &           &       &      &       &      &
&       &      &       &      &       &      \\
 AY Cet   & 12/12/92 & 0.61 & C &     -     & 0.589 & 1.52 & 0.525 & 1.38 &
& 0.711 & 0.79 & 0.657 & 0.81 & -     & -    \\
          &          &      &   &           &       &      &       &      &
&       &      &       &      &       &      \\
 HD 12545 & 15/12/92 & 0.55 & - &     -     & 4.534 & 3.80 & 3.897 & 3.84 &
& 4.874 & 4.36 & 4.308 & 4.34 & 1.243 & 0.81 \\
          &          &      &   &           &       &      &       &      &
&       &      &       &      &       &      \\
 6 Tri    & 15/12/92 & 0.87 & C & 0.20/0.80 & 0.393 & 0.76 & 0.304 & 0.55 &
& 0.566 & 0.38 & 0.476 & 0.36 & -     & -    \\
          &          &      &   &           &       &      &       &      &
&       &      &       &      &       &      \\
 V1149 Ori& 04/03/93 & 0.19 & C &     -     & 1.697 & 2.10 & 1.570 & 1.72 &
& 1.971 & 1.56 & 1.915 & 1.59 & 0.277 & 0.23 \\
 "        & 04/03/93 & 0.19 & C &     -     & 1.667 & 2.00 & 1.562 & 1.64 &
& 1.963 & 1.55 & 1.928 & 1.64 & 0.262 & 0.22 \\
          &          &      &   &           &       &      &       &      &
&       &      &       &      &       &      \\
 RZ Cnc   & 31/01/88 & 0.36 & H &   0.80    & 1.302 & 1.26 & 1.286 & 1.05 &
& 1.427 & 1.11 & 1.323 & 1.21 & 0.195 &      \\
          &          &      & C &   0.20    &       &      &       &      &
& 0.302 & 0.23 & 0.497 & 0.32 & -     & -    \\
 "        & 08/03/93 & 0.44 & H & 0.80/0.20 & 1.619 & 2.12 & 1.499 & 1.90 &
& 1.880 & 1.25 & 1.734 & 1.32 & 0.340 & 0.24 \\
          &          &      &   &           &       &      &       &      &
&       &      &       &      &       &      \\
 DM UMa   & 07/03/93 & 0.85 & - &     -     & 2.733 & 1.95 & 2.254 & 1.69 &
& 3.127 & 2.51 & 2.890 & 2.38 & 0.745 & 0.61 \\
          &          &      &   &           &       &      &       &      &
&       &      &       &      &       &      \\
 DK Dra   & 26/11/86 & 0.44 & T & 0.50/0.50 & 1.253 & 1.78 & 1.140 & 1.38 &
& 1.464 & 1.17 & 1.388 & 1.08 & -     & -    \\
 "        & 29/01/88 & 0.10 & T & 0.50/0.50 & 1.496 & 2.70 & 1.377 & 2.64 &
& 1.614 & 1.41 & 1.489 & 1.37 & -     & -    \\
 "        & 31/01/88 & 0.13 & T & 0.50/0.50 & 1.503 & 2.56 & 1.155 & 1.50 &
& 1.640 & 1.25 & 1.464 & 1.17 & -     & -    \\
 "        & 31/01/88 & 0.13 & T & 0.50/0.50 & 1.298 & 2.01 & 1.312 & 2.05 &
& 1.594 & 1.21 & 1.493 & 1.67 & -     & -    \\
 "        & 07/03/93 & 0.02 & T & 0.50/0.50 & 1.844 & 2.47 & 1.700 & 2.04 &
& 2.007 & 1.72 & 1.924 & 1.71 & 0.208 & 0.19 \\
 "        & 09/03/93 & 0.05 & T & 0.50/0.50 & 1.730 & 2.47 & 1.558 & 1.87 &
& 1.870 & 1.65 & 1.783 & 1.56 & 0.164 & 0.14 \\
          &          &      &   &           &       &      &       &      &
&       &      &       &      &       &      \\
 4 UMi    & 05/03/93 & 0.86 & - &     -     & 0.100 & 0.41 & 0.065 & 0.28 &
& 0.145 & 0.09 & 0.107 & 0.09 & -     & -    \\
          &          &      &   &           &       &      &       &      &
&       &      &       &      &       &      \\
 GX Lib   & 13/07/89 & 0.36 & - &     -     & 0.900 & 1.40 & 0.761 & 0.96 &
& 1.047 & 0.68 & 1.116 & 0.72 & -     & -    \\
 "        & 14/07/89 & 0.44 & - &     -     & 0.619 & 1.30 & 0.538 & 0.81 &
& 0.754 & 0.62 & 0.849 & 0.61 & -     & -    \\
 "        & 05/03/93 & 0.83 & - &     -     & 0.689 & 1.37 & 0.555 & 0.82 &
& 0.799 & 0.61 & 0.801 & 0.59 & -     & -    \\
          &          &      &   &           &       &      &       &      &
&       &      &       &      &       &      \\
 DR Dra   & 13/07/89 & 0.67 & C &     -     & 1.164 & 2.41 & 1.045 & 1.85 &
& 1.320 & 1.32 & 1.296 & 1.40 & 0.155 & 0.15 \\
          & 09/03/93 & 0.09 & C &     -     & 1.445 & 3.11 & 1.098 & 1.87 &
& 1.553 & 1.38 & 1.375 & 1.36 & 0.237 & 0.21 \\
          &          &      &   &           &       &      &       &      &
&       &      &       &      &       &      \\
\noalign{\smallskip}
\hline
\end{tabular}
\end{flushleft}
\end{table*}

\begin{table*}[p]
\caption[]{Ca~{\sc ii} H \& K lines measures in the observed and
subtracted spectrum
(Single stars or components of visual binaries)
\label{tab:medhykwilson} }
\begin{flushleft}
\scriptsize
\begin{tabular}{l l c c c c c c c c c c c c c}
\hline
\noalign{\smallskip}
 &    &  &   & \multicolumn{4}{c}{Reconstruction of the profile} &\ &
\multicolumn{6}{c}{Spectral subtraction} \\
\cline{5-8}\cline{10-15}
\noalign{\smallskip}
\noalign{\smallskip}
 HD  & Name &  \multicolumn{2}{c}{F({\rm 1.0 \AA})} &
EW & C(K) & EW & C(H) & &
EW & I(K) & EW & I(H) & EW & I(H$\epsilon$) \\
 & & (K) & (H) & (K) & & (H) & & & (K) &  & (H) &   & (H$\epsilon$) &  \\
\noalign{\smallskip}
\hline
 {\bf F }  \\
\hline
\noalign{\smallskip}
 13480 B & 6 Tri B       & 0.226 & 0.251 & -     & -    & -     & -    &
& -     & -    & -     & -    & -    & -   \\
 120136& $\tau$ Boo      & 0.148 & 0.181 & -     & -    & -     & -    &
& -     & -    & -     & -    & -    & -   \\
 124850  &$\iota$ Vir    & 0.229 & 0.280 & -     & -    & -     & -    &
& -     & -    & -     & -    & -    & -   \\
 187013  & 17 Cyg        & 0.116 & 0.139 & -     & -    & -     & -    &
& -     & -    & -     & -    & -    & -   \\
 212754  & 34 Peg        & 0.099 & 0.117 & -     & -    & -     & -    &
& -     & -    & -     & -    & -    & -   \\
 216385 & $\sigma$ Peg   & 0.104 & 0.125 & -     & -    & -     & -    &
& -     & -    & -     & -    & -    & -   \\
 187691 & o Aql          & 0.111 & 0.134 & -     & -    & -     & -    &
& -     & -    & -     & -    & -    & -   \\
 142373 & $\chi$ Her     & 0.108 & 0.128 & -     & -    & -     & -    &
& -     & -    & -     & -    & -    & -   \\
 194012 & HR 7793        & 0.163 & 0.184 & -     & -    & -     & -    &
& -     & -    & -     & -    & -    & -   \\
 45067  & HR 2313        & 0.112 & 0.130 & -     & -    & -     & -    &
& -     & -    & -     & -    & -    & -   \\
 6920   & 44 And         & 0.174 & 0.183 & -     & -    & -     & -    &
& -     & -    & -     & -    & -    & -   \\
 107213 & 9 Com          & 0.104 & 0.122 & -     & -    & -     & -    &
& -     & -    & -     & -    & -    & -   \\
 136202 & 5 Ser          & 0.122 & 0.154 & -     & -    & -     & -    &
& -     & -    & -     & -    & -    & -   \\
 154417 & HR 6349        & 0.199 & 0.230 & 0.049 & 0.25 & 0.051 & 0.23 &
& 0.112 & 0.11 & 0.164 & 0.10 & -    & -   \\
 43587  & HR 2251        & 0.119 & 0.144 & -     & -    & -     & -    &
& -     & -    & -     & -    & -    & -   \\
\noalign{\smallskip}
\hline
{\bf G} \\
\hline
\noalign{\smallskip}
115383  & 59 Vir         & 0.267 & 0.306 & 0.099 & 0.41 & 0.100 & 0.39 &
& 0.175 & 0.20 & 0.198 & 0.21 & -    & -   \\
152792  &                & 0.111 & 0.132 & -     & -    & -     & -    &
& -     & -    & -     & -    & -    & -   \\
114710  & $\beta$ Com    & 0.155 & 0.186 & -     & -    & -     & -    &
& -     & -    & -     & -    & -    & -   \\
206860  & HN Peg         & 0.280 & 0.297 & 0.110 & 0.49 & 0.101 & 0.40 &
& 0.262 & 0.24 & 0.254 & 0.23 & -    & -   \\
29645   & HR 1489        & 0.101 & 0.127 & -     & -    & -     & -    &
& -     & -    & -     & -    & -    & -   \\
13974   & $\delta$ Tri   & 0.184 & 0.202 & -     & -    & -     & -    &
& -     & -    & -     & -    & -    & -   \\
98231   & $\xi$ UMa A    & 0.188 & 0.204 & 0.029 & 0.17 & 0.031 & 0.16 &
& 0.046 & 0.09 & 0.059 & 0.08 & -    & -   \\
218739  & ADS 16557 A    & 0.304 & 0.301 & 0.139 & 0.58 & 0.120 & 0.38 &
& 0.224 & 0.26 & 0.121 & 0.22 & -    & -   \\
146362 & $\sigma^{1}$ CrB& 0.233 & 0.262 & 0.037 & 0.16 & 0.028 & 0.12 &
& 0.268 & 0.09 & 0.361 & 0.11 & -    & -   \\
13421   & 64 Cet         & 0.083 & 0.109 & -     & -    & -     & -    &
& -     & -    & -     & -    & -    & -   \\
"       & "              & 0.103 & 0.124 & -     & -    & -     & -    &
& -     & -    & -     & -    & -    & -   \\
190406  & 15 Sge         & 0.145 & 0.166 & -     & -    & -     & -    &
& -     & -    & -     & -    & -    & -   \\
-       & Sol            & 0.160 & 0.187 & -     & -    & -     & -    &
& -     & -    & -     & -    & -    & -   \\
143761  & $\rho$ CrB     & 0.117 & 0.141 & -     & -    & -     & -    &
& -     & -    & -     & -    & -    & -   \\
81809   & HR 3750        & 0.139 & 0.151 & -     & -    & -     & -    &
& -     & -    & -     & -    & -    & -   \\
9562    & HR 448         & 0.095 & 0.127 & -     & -    & -     & -    &
& -     & -    & -     & -    & -    & -   \\
12235   & 112 Psc        & 0.114 & 0.136 & -     & -    & -     & -    &
& -     & -    & -     & -    & -    & -   \\
217014  & 51 Peg         & 0.110 & 0.127 & -     & -    & -     & -    &
& -     & -    & -     & -    & -    & -   \\
20630&$\kappa^{1}$ Cet   & 0.307 & 0.320 & 0.143 & 0.63 & 0.124 & 0.46 &
& 0.258 & 0.25 & 0.307 & 0.22 & -    & -   \\
115617  & 61 Vir         & 0.126 & 0.152 & -     & -    & -     & -    &
& -     & -    & -     & -    & -    & -   \\
190360  & HR 7670        & 0.105 & 0.130 & -     & -    & -     & -    &
& -     & -    & -     & -    & -    & -   \\
182488  & HR 7368        & 0.127 & 0.144 & -     & -    & -     & -    &
& -     & -    & -     & -    & -    & -   \\
131156 A & $\xi$ Boo A   & 0.428 & 0.428 & 0.233 & 0.94 & 0.229 & 0.88 &
& 0.477 & 0.47 & 0.463 & 0.45 & -    & -   \\
144287  &                & 0.129 & 0.153 & -     & -    & -     & -    &
& -     & -    & -     & -    & -    & -   \\
101501  & 61 UMa         & 0.262 & 0.277 & 0.087 & 0.35 & 0.066 & 0.24 &
& 0.121 & 0.16 & 0.098 & 0.31 & -    & -   \\
182572  & 31 Aql         & 0.118 & 0.139 & -     & -    & -     & -    &
& -     & -    & -     & -    & -    & -   \\
188512  & $\beta$ Aql    & 0.106 & 0.126 & -     & -    & -     & -    &
& -     & -    & -     & -    & -    & -   \\
158614  & HR 6516        & 0.127 & 0.169 & -     & -    & -     & -    &
& -     & -    & -     & -    & -    & -   \\
\noalign{\smallskip}
\hline
 {\bf K} \\
\hline
\noalign{\smallskip}
3651    & 54 Psc         & 0.109 & 0.122 & -     & -    & -     & -    &
& -     & -    & -     & -    & -    & -   \\
190404  &                & 0.141 & 0.164 & 0.312 & 0.16 & -     & -    &
& -     & -    & -     & -    & -    & -   \\
10476   & 107 Psc        & 0.130 & 0.155 & -     & -    & -     & -    &
& -     & -    & -     & -    & -    & -   \\
22072   & HR 1085        & 0.109 & 0.132 & -     & -    & -     & -    &
& -     & -    & -     & -    & -    & -   \\
142091  &$\kappa$ CrB    & 0.103 & 0.115 & -     & -    & -     & -    &
& -     & -    & -     & -    & -    & -   \\
22049 &$\epsilon$ Eri    & 0.520 & 0.522 & 0.344 & 1.33 & 0.305 & 1.07 &
& 0.412 & 0.66 & 0.390 & 0.62 & -    & -   \\
4628    & HR 222         & 0.203 & 0.233 & 0.071 & 0.49 & 0.056 & 0.28 &
& 0.072 & 0.14 & 0.094 & 0.10 & -    & -   \\
16160   & HR 753         & 0.216 & 0.222 & 0.073 & 0.46 & 0.045 & 0.21 &
& -     & -    & -     & -    & -    & -   \\
219134  & HR8832         & 0.183 & 0.206 & 0.065 & 0.54 & 0.052 & 0.37 &
& -     & -    & -     & -    & -    & -   \\
115404  &                & 0.474 & 0.489 & 0.289 & 1.41 & 0.270 & 1.01 &
& 0.323 & 0.58 & 0.311 & 0.57 & -    & -   \\
127665  &$\rho$ Boo      & 0.130 & 0.131 & 0.075 & 0.52 & 0.059 & 0.39 &
& -     & -    & -     & -    & -    & -   \\
131156 B & $\xi$ Boo B   & 1.337 & 1.249 & 1.066 & 2.13 & 0.940 & 1.81 &
& -     & -    & -     & -    & 0.105& -   \\
201091 & 61 Cyg A        & 0.659 & 0.655 & 0.453 & 1.93 & 0.381 & 1.38 &
& -     & -    & -     & -    & 0.041& -   \\
201092 & 61 Cyg B        & 1.074 & 1.002 & 0.825 & 2.65 & 0.691 & 2.05 &
& -     & -    & -     & -    & 0.039& -   \\
\noalign{\smallskip}
\hline
\end{tabular}
\end{flushleft}
\end{table*}

\end{document}